\documentclass[11pt]{article}

\usepackage{microtype}%if unwanted, comment out or use option "draft"

\usepackage{bm} % RSADD
\usepackage{cprotect} % RSADD for verbatim in caption

\usepackage{amsfonts,amsmath,amsthm}
\usepackage{graphicx}
\usepackage{listings}
\usepackage{hyperref}
\usepackage{xcolor}
\usepackage[margin=1in]{geometry}

\newcommand{\real}{\mathbb{R}}
\title{Aggregative Coarsening for Multilevel Hypergraph Partitioning\footnote{This material is based upon work supported by the National Science Foundation under Grant No. 1522751.}}
\author{
  Ruslan Shaydulin \\ 
  School of Computing\\ 
  Clemson University \\
    rshaydu@clemson.edu
  \and
  Ilya Safro \\ 
  School of Computing \\ 
  Clemson University\\
  isafro@clemson.edu
}

\begin{document}

\maketitle

\begin{abstract}
Algorithms for many hypergraph problems, including partitioning, utilize multilevel frameworks to achieve a good trade-off between the performance and the quality of results. In this paper we introduce two novel aggregative coarsening schemes and incorporate them within state-of-the-art hypergraph partitioner Zoltan. Our coarsening schemes are inspired by the algebraic multigrid and stable matching approaches. We demonstrate the effectiveness of the developed schemes as a part of multilevel hypergraph partitioning framework on a wide range of problems.
\end{abstract}

\section{Introduction}
Hypergraph is a generalization of graph. Whereas in a graph each edge connects only two vertices, in a hypergraph a hyperedge can connect an arbitrary number of vertices. In many cases this allows hypergraph  to better capture the underlying structure of the problem. 
In graph partitioning (GP), the goal is to split the vertex set of a graph into approximately even parts while minimizing the number of the edges in a cut \cite{buluc2016recent}. Hypegraph partitioning problem (HGP) extends it to hypergraphs. Hypergraph partitioning has many applications in fields ranging from VLSI design~\cite{karypis1999multilevel} to parallel matrix multiplication~\cite{catalyurek1999hypergraph} to classification~\cite{zhou2006learning} to optimizing distributed systems~\cite{kumar2014sword,curino2010schism}, among others~\cite{devine2005new,overlapping-matrix-hyper}. 

Hypergraph partitioning is NP-hard~\cite{garey2002computers} and relies on heuristics in practice. Many state-of-the-art graph and hypergraph partitioners utilize the multilevel approach \cite{buluc2016recent}. In multilevel methods, the original problem is iteratively coarsened by creating a hierarchy of smaller problems, until it becomes small enough to be solved. Then the coarsest problem is solved and the solution is iteratively projected onto finer levels and refined. 
Multilevel algorithms for HGP are typically generalizations of multilevel algorithms for graph partitioning, those in turn drawing inspiration from multigrid and other multiscale optimization techniques~\cite{brandt2003multigrid}. Hypergraph partitioning is less well-studied than graph partitioning~\cite{curino2010schism} and there is a relative lack of advanced coarsening schemes compared to GP. 

Our \emph{main contribution} are two novel aggregative coarsening schemes for HGP that are inspired by algebraic multigrid and stable matching methods. We expand and build on the insights from an unfinished attempt to build a coarsening scheme for HGP using algebraic multigrid ideas, published in Sandia Labs Summer Reports~\cite{bulucc2008towards}. At each coarsening level we split the set of vertices into the set of seeds and the set of non-seeds. Each seed becomes a center of an aggregate which will, in turn, create a node at the next, coarser level. Aggregation rules are established to specify which aggregate a non-seed joins. We investigate two approaches to establishing aggregation rules. One approach is algebraic multigrid-based generatlization of an inner-product matching similar to the matching scheme used in Zoltan\cite{catalyurek2009repartitioning} and PaToH\cite{catalyurek1999hypergraph}. Another approach is inspired by stable matching. Both approaches take advantage of the algebraic distance on hypergraphs when making coarsening decisions. Algebraic distance is a vertex similarity measure that extends simple measures such as hyperedge weights to better capture the structure of the hypergraphs~\cite{shaydulin2017relaxation}. While we outperform existing solvers on many instances, it is clear that final performance of HGP solvers heavily depends on the refinement. It is not the goal of this paper to outperform all existing HGP solvers. Instead, we would like to demonstrate that given similar uncoarsening  schemes, the proposed coarsening schemes are at least as beneficial as traditional matching-based approaches. 

\section{Preliminaries}

A hypergraph is an ordered pair of sets $(V,E)$, where $V$ is the set of vertices and $E$ is the set of hyperedges. Each hyperedge $e\in E$ is a nonempty subset of $V$. In this paper we make use of a graph representation of a hypergraph called "star expansion". Star expansion graph $(V',E')$ of a hypergraph $(V,E)$ is an undirected bipartite graph with hypergraph vertices $V$ on one side, hyperedges $E$ on another and edges connecting hyperedges with the vertices they contain. Concretely, $V' = V\cup E$, $E' = \{(v,e)\mid e\in E, v\in e\subset V\}$. We will be referring to hyperedges as simply edges where it does not cause confusion. 
 Both vertices and edges of the hypergraph are positively weighted. By $w(v)$ and $w(e)$, we denote weighting functions for nodes and edges, where $v\in V$ and $e\in E$. For both nodes and edges,  a weight of zero practically means that corresponding nodes or edges do not exist (or do not affect the optimization and solution).

\subsection{Hypergraph partitioning}

In hypergraph k-partitioning the goal is to split the set of vertices $V$ into $k$ disjoint subsets $(V_1, \ldots, V_k)$ such that a metric on the cut is minimized subject to imbalance constraint. Here the cut is defined as the set of edges that span more than one partition, i.e., 
\begin{equation}
E_{\text{cut}} = \{e\in E \mid \exists i\neq j \text{ and } k\neq l \text{ for which }v_i, v_j\in e, v_i\in V_k \text{ and } v_j \in V_l\}.
\end{equation}

There are multiple ways to define imbalance constraint. We will follow the definition used by the developers of state-of-the-art hypergraph partitioner Zoltan~\cite{devine2006parallel}. The imbalance is therefore defined as the ratio between the total weight of vertices in the largest partition and the average sum of weights of vertices over all partitions. We define the imbalance as 
\begin{equation}
imbal=\frac{\sum_{v\in V_{\text{max}}}w(v)}{\frac{1}{k}\sum_{v\in V}w(v)},
\end{equation}
% \noindent
where $V_{max}$ is the largest partition by weight (i.e., $ \sum_{v\in V_{\text{max}}}w(v) = \max_i\left(\sum_{v\in V_{i}}w(v)\right)$). Imbalance constraint imposes a limit on the value of $imbal$, e.g.,  imbalance of $5\%$ means $imbal < 1.05$. The cut metric used in this paper is simply total weight of the cut edges, namely, $\sum_{e\in E_{\text{cut}}}w(e)$.

\subsection{Multilevel method}\label{sec:prereq_multilevel}

\begin{figure}[!ht]
  \includegraphics[width=\textwidth]{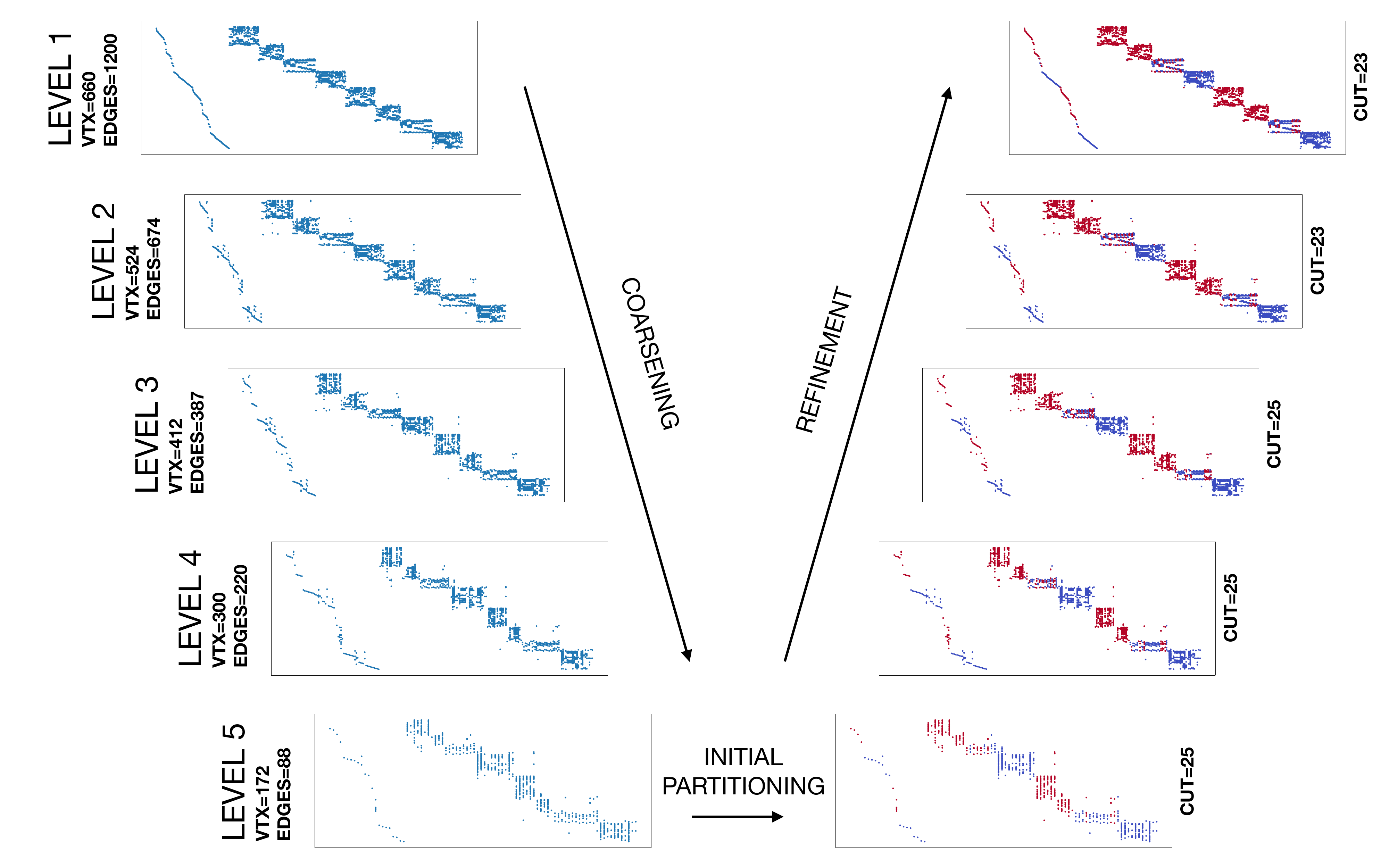}
  \cprotect\caption{Multilevel partitioning of a hypergraph constructed from \verb+LPnetlib/lp_scfxm2+ matrix from SuiteSparse Matrix Collection~\cite{davis2011university} using row-net model: each column becomes a vertex and each row becomes a hyperedge. On the left side of the "V" the hypergraph (represented here as the sparsity pattern of the underlying matrix) is iteratively coarsened. At the bottom of the "V" the hypergraph is partitioned into two parts. This is represented by coloring the columns corresponding to vertices from one part into blue and another into red. On the right side of the "V" the hypergraph is uncoarsened and the partitioning is refined.}\label{fig:v_cycle}
\end{figure}

The main objective of multilevel methods  is to construct a hierarchy of problems, each approximating the original problem but with fewer degrees of freedom. This is achieved by introducing a chain of successive restrictions of the problem domain into low-dimensional or smaller-size domains (coarsening) and solving the coarse problems in them using local processing (uncoarsening) \cite{safro:relaxml}. The coarsening-uncoarsening pipeline is often referred to as V-cycle. The multilevel frameworks combine solutions obtained by the local processing at different levels of coarseness into one global solution.
Typically, for combinatorial optimization problems, the multilevel algorithms are  suboptimal metaheuristics~\cite{Walshaw2004} that incorporate other methods as refinement at all levels of coarseness. Except partitioning, examples can be found in linear ordering \cite{Hu:wavefront,SafroRB06,Safro2006a,SafroRB08}, clustering and community detection \cite{rotta2011multilevel,Blondel08}, and traveling salesman problems \cite{walshaw2002multilevel}. In (H)GP, these algorithms were initially introduced to speed up existing algorithms~\cite{barnard1994fast} but later proved to improve the quality of the solution~\cite{hendrickson1995multi,karypis1998fast}.
A multilevel hypergraph partitioning of a hypergraph constructed from \verb+LPnetlib/lp_scfxm2+ matrix is presented in Figure~\ref{fig:v_cycle}.

During the coarsening stage, for a hypergraph $H=(V,E)$ a hierarchy of decreasing in size hypergraphs $H^0=(V^0,E^0), \ldots, H^l=(V^l,E^l)$ is constructed. Here $l$ denotes the number of levels in multilevel hierarchy. During the initial partitioning stage, the coarsest hypergraph $H^l=(V^l,E^l)$ is partitioned. Finally, during the refinement stage the solution from coarser levels is projected onto finer ones and refined, typically using a local search heuristic.

\subsection{Algebraic distance}

Algebraic distance for hypergraphs is a relaxation-based vertex similarity measure~\cite{shaydulin2017relaxation}. It extends the algebraic distance for graphs~\cite{chen2011algebraic,safro:relaxml,safro:spars} by taking into account the non-pairwise nature of the connections between vertices in a hypergraph. Algebraic distance improves on simpler similarity metrics, such as hyperedge weights, by incorporating information about more distant vertex neighborhood, thus better capturing vertex's place in the global structure of the hypergraph. 
Algebraic distance is inspired by iterative techniques for solving linear systems. An iterative method can be represented in a standard form:
\begin{equation}
x^{(i)} = Hx^{(i-1)}\qquad i = 1,2,3\ldots
\end{equation}
where $H$ is the iteration matrix.

Similarly, algebraic distances are computed at each coarsening level using the following stationary iterative relaxation. Let $A\in R^{|E|\times |V|}$ be hypergraph incidence matrix, i.e., $A_{ij} = 1$ if the hyperedge $i$ contains the vertex $j$. Let $S^v\in\real^{|V|\times|V|}$ and $S^h\in\real^{|E|\times|E|}$ be diagonal matrices such that
\begin{equation}
S_{jj}^v=w(v_j) \quad\text{and}\quad S_{ii}^h=\frac{w(h_i)}{|h_i|},
\end{equation}
where $|h_i|$ denotes the cardinality of the $i$th hyperedge. Denote 
\begin{equation}
W=
\begin{bmatrix}
0 & A^TS^{h} \\
AS^{v} & 0
\end{bmatrix} 
\end{equation}
and let $D$ be the diagonal matrix with elements $D_{jj}=\sum_iW_{ij}$. Then the iterative step is defined as

\begin{equation}
x^{(i)}=\frac{1}{r-l}\Big[\underbrace{\omega D^{-1}Wx^{(i-1)}+(1-\omega)x^{(i-1)}}_{{x^{*}}^{(i-1)}}\Big]
-\frac{r+l}{2(r-l)}\bm{1}
\end{equation}
where $\bm{1}$ is the vector of all ones, and $r$ and $l$ are the maximum and the minimum of the elements in ${x^{*}}^{(i-1)}$, respectively. 
We can simplify the update formula as
\begin{equation}\label{eq:xk}
x^{(i)}=\alpha^{(i-1)}Hx^{(i-1)}+\beta^{(i-1)}\bm{1},
\end{equation}
where
\begin{equation}
H=\omega D^{-1}W+(1-\omega)I,\quad
\alpha^{(i-1)}=\frac{1}{r-l},\quad\text{and}\quad
\beta^{(i-1)}=-\frac{r+l}{2(r-l)}.
\end{equation}

The iterative scheme is performed multiple times for different random initial values $x^{(0)}_0, \ldots x^{(0)}_R$ (called test vectors in algebraic multigrid \cite{livne2012lean}) . Then, the algebraic distance between vertices $i$ and $j$ is set to be the maximum over $R$ random initializationsm namely,  $\text{algdist}_{ij}=\max_R |x_i - x_j|$.
For the detailed discussion of algebraic distances on hypergraphs and for convergence analysis of the described iterative scheme the reader is referred to~\cite{shaydulin2017relaxation}.

\section{Related work}\label{sec:related}
Practical approaches to solving HGP typically rely on heuristics. Many have been developed over the years, but the most common approach is multilevel. It is implemented in many state-of-the-art hypergraph partitioners including but not limited to Zoltan~\cite{devine2006parallel}, hMetis~\cite{karypis1999multilevel}, KaHyPar~\cite{shhmss2016alenex} and PaToH~\cite{catalyurek1999hypergraph}. In this section we will briefly describe the multilevel approach used by those state-of-the-art partitioners and discuss existing advanced coarsening schemes for hypergraphs. For a more detailed review of hypergraph partitioning, the reader is referred to~\cite{alpert1995recent,bader2013graph,papa2007hypergraph,trifunovic2006parallel}.

\subsection{Brief overview of multilevel hypergraph partitioning}

\begin{figure}
\centering
  \includegraphics[width=0.8\textwidth]{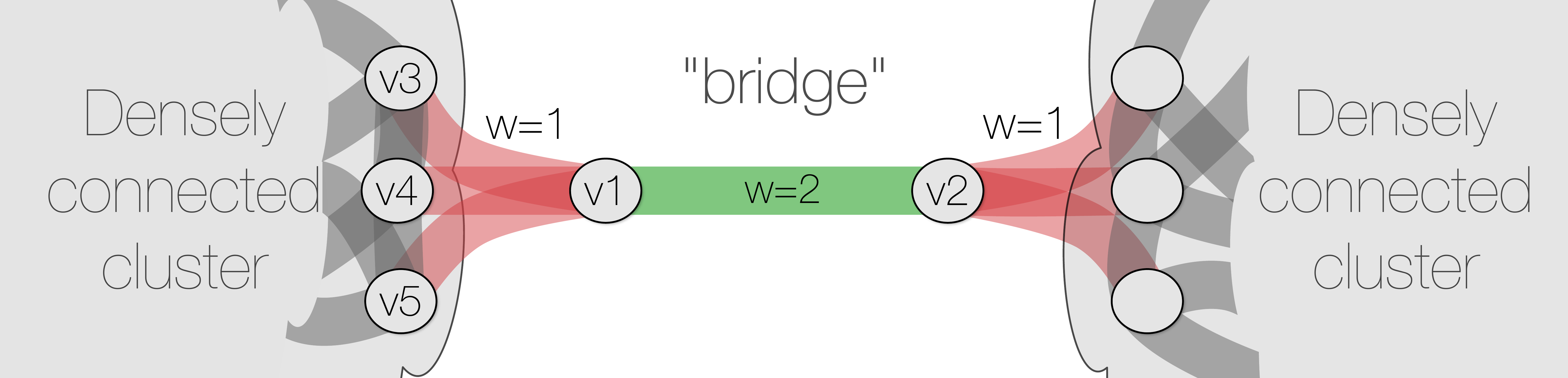}
  \caption{ An example demonstrating the limitations of schemes based on edge weights. Here the "bridge" edge (green) connecting v1 and v2 has weight two and all other edges (red) have unitary weights. The best cut is achieved by cutting the green "bridge" between v1 and v2 and therefore matching v1 with one of the vertices on the left (v3,v4 or v5). However, matching schemes based on edge weights can match v1 with v2 instead and increase the cut.}\label{fig:structure}
\end{figure}

The HGP multilevel frameworks consist of three stages: coarsening, initial partitioning and uncoarsening with refinement. During the coarsening, the hypergraph is approximated via a series of decreasing in size hypergraphs. At each coarsening step, the next hypergraph is formed by matching a group of vertices into one, such that a set of vertices at level $k$ becomes one vertex at level $k+1$. The decision as to which vertices to match is made based on similarity metrics such as inner product (i.e., the total weight of hyperedges connecting two vertices). However, simple metrics often result in a decision that ignores the structure of the hypergraph. Consider the example in Figure~\ref{fig:structure}. It shows two densely connected clusters, separated by a "bridge" between vertices v1 and v2. Schemes that use naive similarity metrics like hyperedge weights might match v1 with v2, whereas an algorithm that considers larger neighborhoods to minimize a cut would prefer to match v1 with either v3, v4 or v5 instead. This example demonstrates the challenges of capturing the hypergraph structure by using only local information. In the refinement stage all the aforementioned state-of-the-art partitioners use a combination of Fiduccia-Mattheyses~\cite{fiduccia1988linear} or Kernighan-Lin~\cite{kernighan1970efficient} with the exception of KaHyPar, which uses a novel localized adaptive local search heuristic~\cite{ahss2017alenex}.

\subsection{Aggregative coarsening}

The standard approach to coarsening used in most state-of-the-art hypergraph partitioners is matching-based. Originally, this meant that at each level pairs of adjacent vertices are selected to become one vertex at the next level. This technique has later been extended to include non-pairwise matchings (i.e., more than two fine vertices can form a coarse vertex). One of the alternative approaches is  aggregative coarsening inspired by algebraic multigrid. In aggregative coarsening, the set of vertices $V$ is separated into disjoint sets of seeds and non-seeds, namely, $C$ and $F$ such that $F\cup C=V$. The non-seed vertices aggregate themselves around the seeds (hence the name aggregative coarsening). The aggregation can be strict ($F$-vertices are not split) and weighted ($F$-vertices can be split between multiple seeds with vertex weight conservation). At the refinement stage, the partitioning decision (i.e., partition assignment) is interpolated from each seed to the non-seeds in its aggregate. This separation between seed and non-seeds helps to introduce additional guarantees. For example, on graphs Safro et al.~\cite{safro2006graph} introduce the notion of strong connection and guarantee that each vertex in the graph is strongly connected to at least one seed. The weighted aggregation was initially introduced for several cut problems on graphs \cite{SafroT11,safro:relaxml,SafroRB08} including GP \cite{amg-sss12}. There was an unfinished attempt to extend this approach to hypergraphs. Bulu\c{c} and Boman~\cite{bulucc2008towards} describe several challenges in applying aggregative coarsening to hypergraphs, as well as propose two very similar coarsening schemes, strict and weighted. In this paper, we limit our discussion to strict aggregation.

In aggregative coarsening, two main questions have to be addressed: seed selection and aggregation of non-seeds around seeds. In the process of seed selection, Bulu\c{c} and Boman~\cite{bulucc2008towards} follow Safro et al.~\cite{safro2006graph} in using the concept of \emph{future volumes}. Future volume is a measure of how many vertices a seed can incorporate into itself (in other words, how large a vertex can grow). They propose computing future volumes on the star expansion of the hypergraph (thus limiting the complexity), then iteratively adding vertices with high future volumes to the set of seeds $C$ until $|C|$ reaches a certain threshold. Aggregation rules are established on the star expansion of the hypergraph. Seeds and non-seeds select a constant number of adjacent hyperedges to "invade" based on the exclusive coarseness (a metric indicating how many seeds an hyperedge contains). 

\section{Two Aggregation Algorithms}

Our algorithm combines the ideas of aggregative coarsening described in~\cite{safro2006graph} and~\cite{bulucc2008towards} with the~algebraic distance~\cite{safro:relaxml,shaydulin2017relaxation}. Aggregative coarsening is a two-step process, so we have to~address both the seed selection and the rules of aggregation. At each coarsening level, a set of seeds is selected and each seed is assigned a set of non-seeds to form a cluster. The cluster at a given coarsening level becomes one vertex at the next level.

Both introduced schemes utilize algebraic distances by augmenting hyperedge weights with algebraic weights. We define the algebraic weight of hyperedge $e$  as an  inverse of the algebraic distance between two farthest apart vertices in $e$, i.e.,
\begin{equation}
\rho(e) = 1\ /\ \max_{i,j\in e}\text{algdist}_{ij}. 
\end{equation}

\subsection{Seed selection}

For the seed selection we utilize two core concepts: \textit{future volumes} and \textit{strong connection}. The main goal is to construct a set of seeds $C$ such that every vertex in the graph is \textit{strongly connected} to $C$. 
We define \textit{strong connection} as follows: the vertex $i\in F$ is \textit{strongly connected} to $C$ if the sum of algebraic weights of the edges connecting it to $C$ is more than a certain fraction of the total algebraic weight of incident edges:
\begin{equation}
i \text{ is strongly connected to }C \qquad \iff \qquad \frac{\Sigma_{j\in C}\rho(e_{ij})}{\Sigma_{j}\rho(e_{ij})} > Q,
\end{equation}
where $Q$ is a parameter (in our experiments $Q=0.5$). 
The \textit{future volume} of a vertex is a measure of how large an aggregate seeded by it can grow. Intuitively, we want to add the vertices with very high volume (or the ones that might become centers of the aggregates of very high volume) to the set of seeds. \textit{Future volume} of a vertex is defined as follows (note that here we use the hyperedge weights $w$ and not the algebraic weights $\rho$):
\begin{equation}\label{eq:fv}
fv(i) = w(i) + \Sigma_{j}w(j)\frac{w(e_{ij})}{\Sigma_k w(e_{jk})}.
\end{equation}

We begin the construction of set $C$ by computing future volumes for all vertices. Then, we initialize $C$ with vertices with large future volumes (if mean future volume is $m_{fv}$ and standard deviation of the distribution of future volumes is $\sigma_{fv}$, then $i\in C \iff fv(i) > m_{fv}+2\sigma_{fv})$ and initialize $F$ with all other vertices, such that $F\cup C = V$. After that the future volumes of vertices in $F$ are recomputed, only taking into account connections with other vertices in $F$ (i.e., in Equation~(\ref{eq:fv}) assume  $w(e_{ij}) = 0$ if $j\in C$ or $i\in C$). Finally, vertices in $F$ are visited in order of decreasing future volume and added to the set $C$ if they are not strongly connected to $C$. Note that at the end of this process each vertex in $V$ is strongly connected to the set $C$ and $F\cup C=V$. Pseudocode for this procedure is presented in Listing~\ref{code:seed_sel}.

\begin{lstlisting}[caption={Seed selection},label=code:seed_sel,captionpos=t,float,abovecaptionskip=-\medskipamount,mathescape=true]
for i in V:
	fv[i] = w[i] + $\Sigma_{j}$w[j](w[$e_{ij}$] / $\Sigma_k$w[$e_{jk}$])
for i in V:
	if fv[i] > mean(fv) + 2 * stdev(fv):
    	C.insert(i)
    else:
    	F.remove(i)
for i in F:
	fv[i] = w[i] + $\Sigma_{j\in F}$w[j](w[$e_{ij}$] / $\Sigma_{k\in F}$w[$e_{jk}$])
    
for i in sort_indices(fv):
	if $\Sigma_{j\in C}$w[$e_{ij}$] / $\Sigma_{j}$w[$e_{ij}$] < Q:
		C.insert(i)
		F.remove(i)
\end{lstlisting}

\subsection{Aggregation}

We investigate two approaches to establishing the rules of aggregation. First approach is a scheme similar to inner-product matching used in Zoltan\cite{catalyurek2009repartitioning} and PaToH\cite{catalyurek1999hypergraph} but applied in algebraic multigrid setting. Second approach consists of computing a stable assignment~\cite{gusfield1989stable} between vertices of $C$ and $F$. Both approaches take advantage of algebraic distances as a similarity measure when establishing aggregation rules. 
% The stable assignment approach is more computationally intensive, but tends to produce better results (see Section~\ref{sec:results}).

% \added[id=is]{do you add epsilon to (11)?}
Inner-product aggregation proceeds by visiting the non-seed vertices in the random order. For each unmatched vertex $v\in F$, a neighboring seed $u\in C$ with the highest inner product is selected and $v$ is added to the cluster $C_u$ seeded by it. The inner product is defined as the total algebraic weight of the edges connecting $v$ with the seed $u$. Concretely, $\text{ipm}(v,u)=\Sigma_{e \mid v,u\in e}\ \rho(e)$. See Listing~\ref{code:ipm_match} for pseudocode. We experimented with visiting the non-seeds in order of decreasing future volume and with using connectivity to make decisions when establishing aggregation rules. These approaches are more computationally intensive and do not produce better results (see Appendix~\ref{app:comparison} for the results). 

\begin{lstlisting}[caption={Inner-product aggregation},label=code:ipm_match,captionpos=t,float,abovecaptionskip=-\medskipamount,mathescape=true,belowskip=-0.8 \baselineskip]
for i in F:
	j = argmax$_{u\in C}$ ipm(v,u)
	$C_\text{j}$.insert(i)
\end{lstlisting}

Stable assignment aggregation begins by constructing preference lists. Each seed orders adjacent non-seeds in the order of decreasing total algebraic weight of the hyperedges connecting them (and vice versa): $\text{pref}_i(j) = \Sigma \rho(e_{ij})$. Then the stable assignment is computed using an algorithm similar to the classical one described in~\cite{gale1962college}. Each seed in $C$ proposes to non-seeds in its preference list. If the non-seed does not have a better offer, it tentatively accepts the proposal and is put on the waitlist. If that non-seed later receives a better offer (i.e., an offer from a seed that ranks higher on its preference list), it rejects the current offer and the rejected seed proposes to the next candidate on its preference list. To discourage the creation of very large clusters, we limit the size of waitlist for a seed to the maximal vertex weight on a given coarsening level times three plus ten: $\text{len(waitlist)} = 3\times\text{max\_vtx\_wgt}+10$. Procedure terminates when each non-seed has been assigned to a waitlist or a seed has been rejected by every non-seed. At this point each seed forms a cluster with all vertices on its waitlist, subject to size constraint (we guarantee that no cluster can be larger than total vertex weight over the number of parts). The fact that we use a classical problem as a subproblem in our heuristic allows us to potentially leverage the previous work in optimizing and parallelizing stable assignment, such as~\cite{lu2003parallel},\cite{manne2016stable} and \cite{georgiadis2013overlays}. The pseudocode is presented in Listing~\ref{code:stb_match}.

\begin{lstlisting}[caption={Stable matching aggregation},label=code:stb_match,captionpos=t,float,abovecaptionskip=-\medskipamount,mathescape=true, belowskip=-\baselineskip]
def propose(i):
	for j in pref_list[i]:
		if waitlist[i].size > threshold:
			return
		if propos[j] == -1: // j holds no proposal
			propos[j] = i
			waitlist[i].push_back(j)
			continue
		if i is preferable to propos[j]:
			rejected = propos[j]
			propos[j] = i
			waitlist[i].push_back(j)
			propose(rejected)

// Step 1: compute preference lists
for i in F:
	for j in seed_neighbors[i]:
		pref[j] = $\Sigma \rho(e_{\text{ij}})$  // pref is a hashtable
	for j in sort_by_value(pref).keys():
		pref_list[i].push_back(j)
for i in C:
	for j in non_seed_neighbors[i]:
		pref[j] = $\Sigma \rho(e_{\text{ij}})$
	for j in sort_by_value(pref).keys():
		pref_list[i].push_back(j)

// Step 2: compute stable assignment
for i in C:
	propose(i)
\end{lstlisting}

\section{Results}\label{sec:results}

We implemented all algorithms described in this paper within Zoltan~\cite{devine2006parallel} package of the Trilinos Project~\cite{heroux2005overview}. Zoltan is an open-source toolkit of parallel combinatorial scientific computing algorithms~\cite{devine2006parallel}. It includes a hypergraph partitioning algorithm PHG (Parallel HyperGraph partitioner) and interfaces to PaToH and hMetis2. We added our new coarsening schemes and left other phases of the multilevel framework unchanged. Our implementation, data, and full results are available at \url{http://bit.ly/aggregative2018code}. 

The hypergraphs in our benchmark are generated from a selection of matrices using the row-net model. In the row-net model, each column of the matrix represents a vertex, each row represents an edge and a vertex $j$ belongs to the hyperedge $i$ if there is a non-zero element at the intersection of $j$-th column and $i$-th row, i.e., $A_{ij}\neq 0$. All  matrices (more than 300) were obtained from SuiteSparse Matrix Collection~\cite{davis2011university} that includes other collections. For each combination of hypergraph/algorithm/set of parameters, we executed 20 experiments.% including . Many of the matrices in the benchmark are derived from linear programming problems.

We compare our algorithm with four state-of-the-art partitioners: hMetis2~\cite{karypis1999multilevel}, PaToH v3.2~\cite{ccatalyurek2011patoh}, Zoltan PHG~\cite{devine2006parallel} and Zoltan-AlgD~\cite{shaydulin2017relaxation}. PaToH is used as a plug-in for Zoltan with default parameters described in Zoltan's User Guide~\cite{zoltanuserguide}. hMetis2 is used in k-way mode with all parameters set to default: greedy first-choice scheme for coarsening, random k-way refinement, and min-cut objective function. The reason we run hMetis in k-way mode is the way hMetis specifies imbalance constraint. In recursive bisection mode, the imbalance constraint is applied \emph{at each bisection step}, therefore relaxing the constraint as the number of parts increases. We found it almost impossible to compare hMetis in recursive bisection mode fairly (i.e., with the same imbalance) with other partitioners. Both Zoltan and PaToH are used in serial mode.

Optimizing the constants in the running time of the proposed algorithms is beyond the scope of this paper. Currently, for the existing unoptimized implementation the running time of  other state-of-the-art hypergraph partitioners is not improved except for those that generate less levels in the hierarchy. In the experiments, the runtime of unoptimized implementation of our algorithms is up to an order of magnitude larger than the runtime of other state-of-the-art partitioners in worst cases. However, we must point out that our algorithm utilizes the building blocks and ideas of algebraic multigrid, which makes it possible to improve the runtime drastically by leveraging a plethora of existing research in optimizing and parallelizing algebraic multigrid solvers(e.g.~\cite{yang2002boomeramg}, \cite{park2015high}). Similarly, there exists extensive research into optimizing the performance of stable matching solvers. Manne et al.~\cite{manne2016stable} demonstrate the connection between graph matchings and stable marriage and show the scalability of Gale-Shapely type algorithms. Munera et al.~\cite{munera2015solving} present an adaptive search formulation of stable marriage problem and take advantage of a Cooperative Parallel Local Search framework~\cite{munera2014parametric}, achieving superlinear speedup. Gelain et al.~\cite{gelain2013local} demonstrate a different efficient local search method for stable marriage problem. 

\begin{figure}
\centering
  \includegraphics[width=\textwidth]{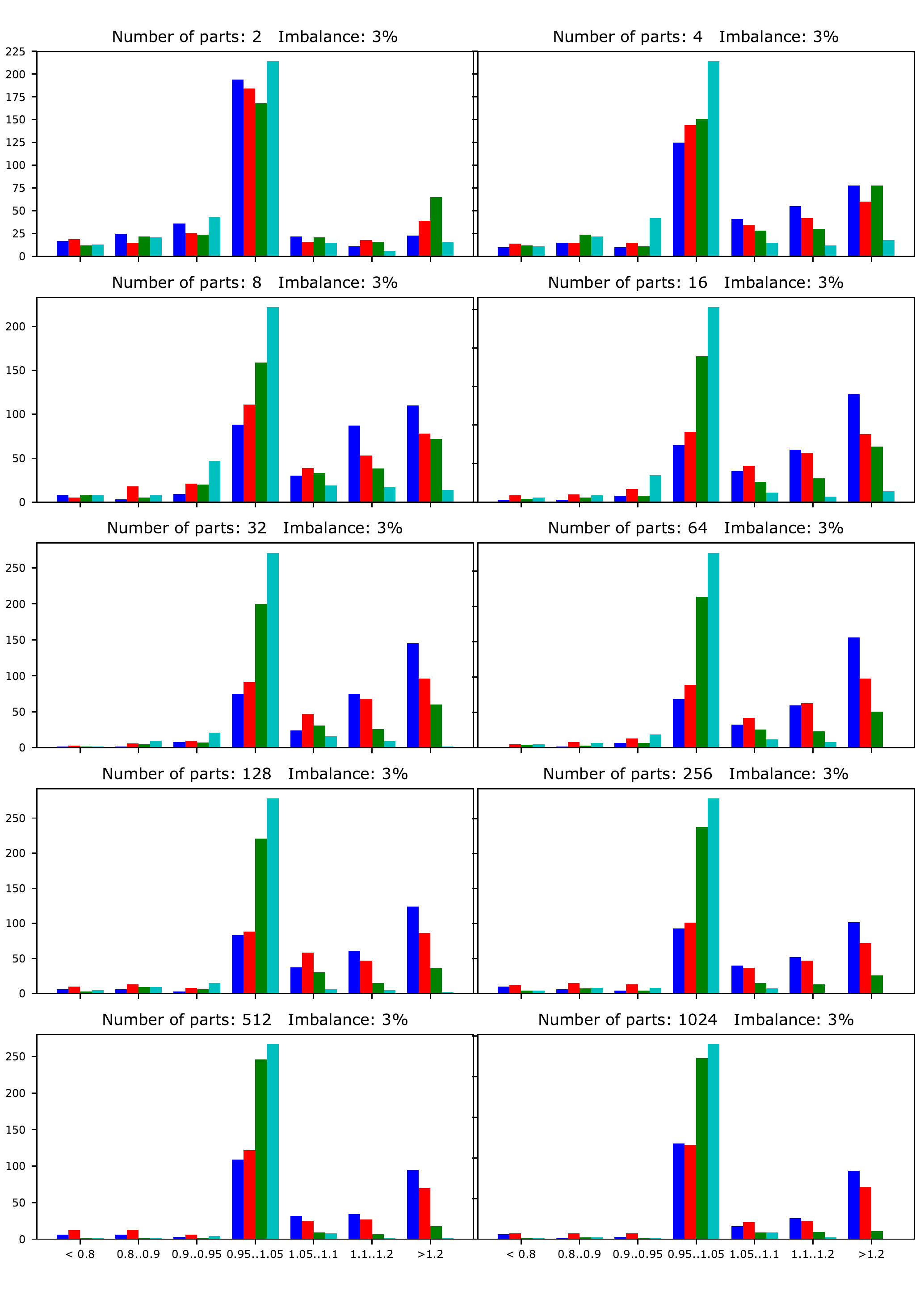}
  \caption{Histogram of $\zeta$ for coarsening using algebraic multigrid inner-product aggregation. Blue rectangle corresponds to PaToH, red to hMetis2, green to Zoltan PHG and cyan to Zoltan-AlgD}\label{fig:aggres}
\end{figure}

\begin{figure}
\centering
  \includegraphics[width=\textwidth]{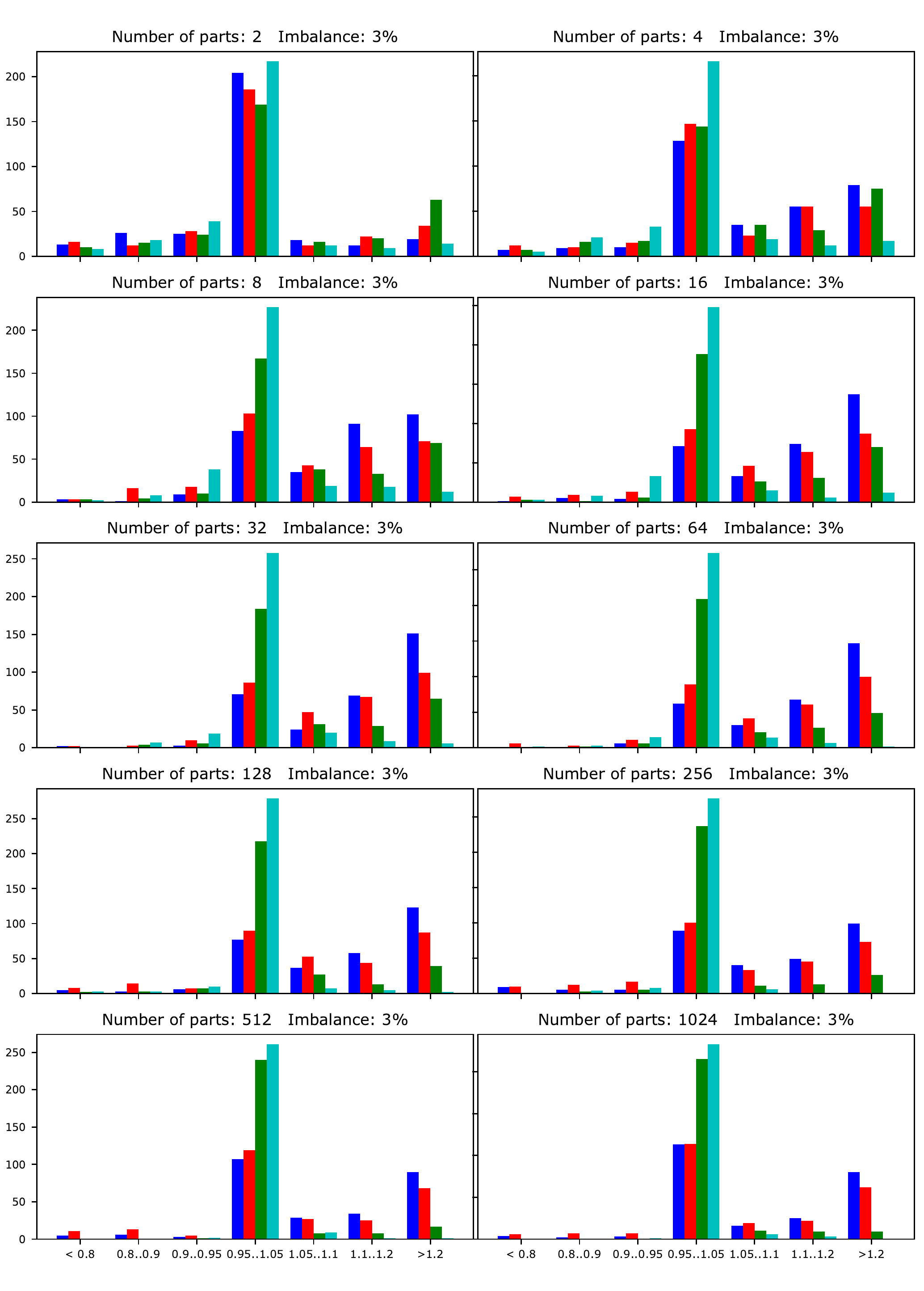}
  \caption{Histogram of $\zeta$ for coarsening using stable matching aggregation. Blue rectangle corresponds to PaToH, red to hMetis2, green to Zoltan PHG and cyan to Zoltan-AlgD}\label{fig:stbres}
\end{figure}

In Figures~\ref{fig:aggres} and~\ref{fig:stbres} the results are presented graphically. In the main body of the paper, we only plot the results for 10\% imbalance. For results for other imbalance factors please refer to Appendix~\ref{app:add_imbal}. In Figure~\ref{fig:aggres}, we show the results of inner-product algebraic multirgid aggregation coarsening. In Figure~\ref{fig:stbres}, the stable matching aggregation is demonstrated. We use frequency histograms to present the distribution of cut differences between our methods and other state-of-the-art hypergraph partitioners. The value being represented (see horizontal axes) is the ratio

\begin{equation}
\zeta = \frac{\text{cut obtained using another partitioner}}{\text{cut obtained using our method}}.
\end{equation}

Each bin corresponds to a range of the ratios (for example, the middle bin corresponds to the differences of less than $\pm 5\%$ and the rightmost to the improvements of $>20\%$). Each rectangle corresponds to a partitioner: blue corresponds to PaToH, red corresponds to hMetis2, green corresponds to Zoltan PHG and cyan corresponds to Zoltan-AlgD. For the full results, please refer to \url{http://bit.ly/aggregative2018results}

The results demonstrate that given the same refinement, the proposed schemes are at least as effective as traditional matching-based schemes, while outperforming them on many instances. Both proposed coarsening schemes almost equally succeed in improving the quality of solvers (see Appendix~\ref{app:add_imbal} for comparison of the performance of two algorithms). Further investigation of the difference between them is a very interesting future research direction, because, in fact, they represent very different algorithms. Since Zoltan utilizes recursive bisectioning scheme, we can see that improvements decrease as number of parts increases. This can be attributed to refinement becoming more and more important as number of parts increases.

\section{Conclusion}

We have presented two novel aggregative coarsening schemes for hypergraphs. The introduced schemes incorporate ideas of algebraic multigrid and stable matching into multilevel hypergraph partitioning framework. We have implemented the described algorithms within state-of-the-art hypergraph partitioner Zoltan and compared their performance against a number of other state-of-the-art partitioners on a large benchmark.

The experimental results demonstrate that given the same uncoarsening, the proposed coarsening schemes perform at least as well as traditional matching-based schemes, while outperforming them on many instances. This suggests that algebraic-multigrid-inspired coarsening schemes have great potential when combined with appropriate refinement.

\newpage

\appendix

\section{Additional results}\label{app:add_imbal}

\begin{figure}
\centering
  \includegraphics[width=\textwidth]{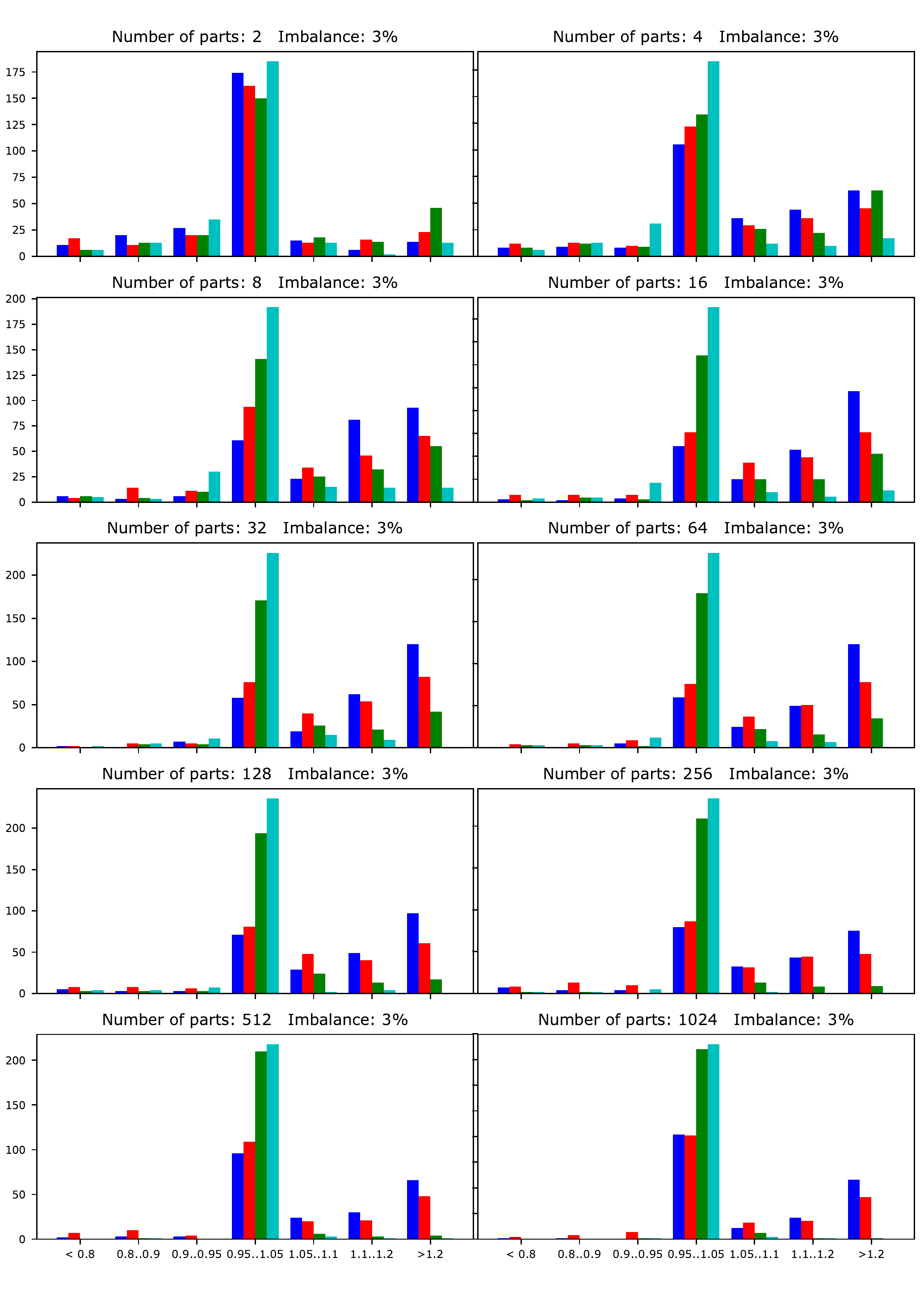}
  \caption{Histogram of $\zeta$ for coarsening using inner-product aggregation with imbalance factor 3\%. Blue rectangle corresponds to PaToH, red to hMetis2, green to Zoltan PHG and cyan to Zoltan-AlgD}\label{fig:aggres103}
\end{figure}

\begin{figure}
\centering
  \includegraphics[width=\textwidth]{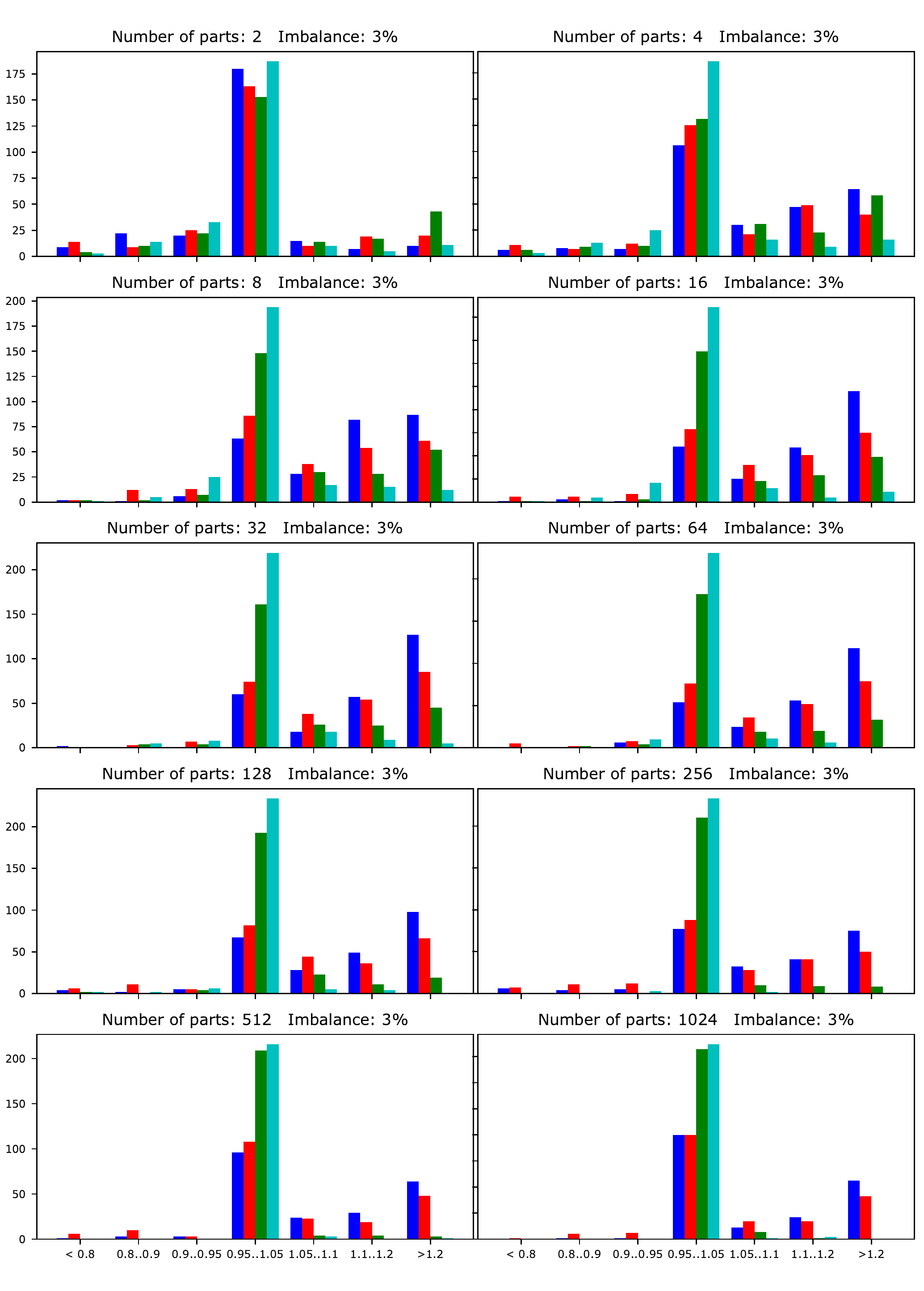}
  \caption{Histogram of $\zeta$ for coarsening using stable matching aggregation with imbalance factor 3\%. Blue rectangle corresponds to PaToH, red to hMetis2, green to Zoltan PHG and cyan to Zoltan-AlgD}\label{fig:stbres103}
\end{figure}

\begin{figure}
\centering
  \includegraphics[width=\textwidth]{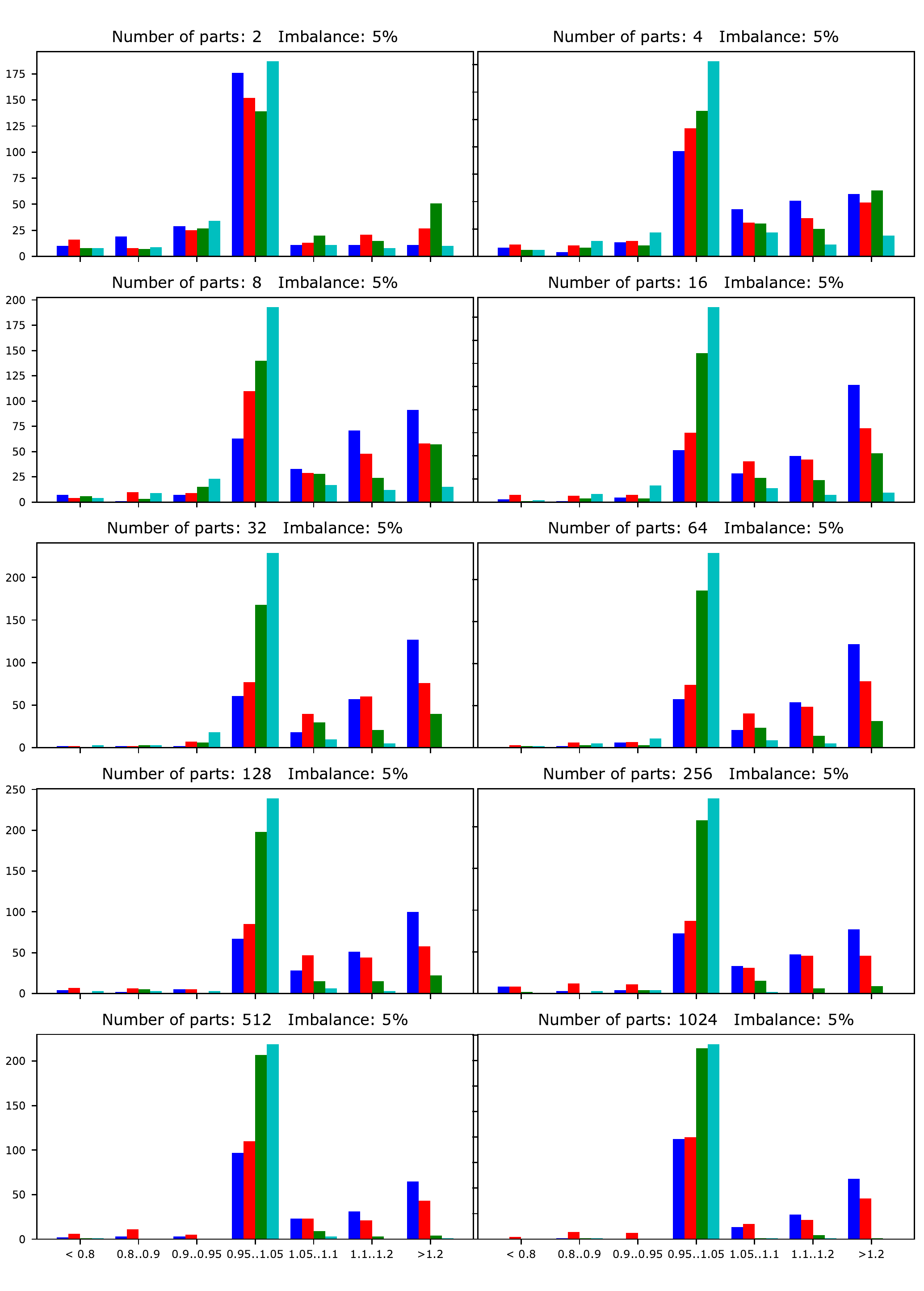}
  \caption{Histogram of $\zeta$ for coarsening using inner-product aggregation with imbalance factor 5\%. Blue rectangle corresponds to PaToH, red to hMetis2, green to Zoltan PHG and cyan to Zoltan-AlgD}\label{fig:aggres105}
\end{figure}

\begin{figure}
\centering
  \includegraphics[width=\textwidth]{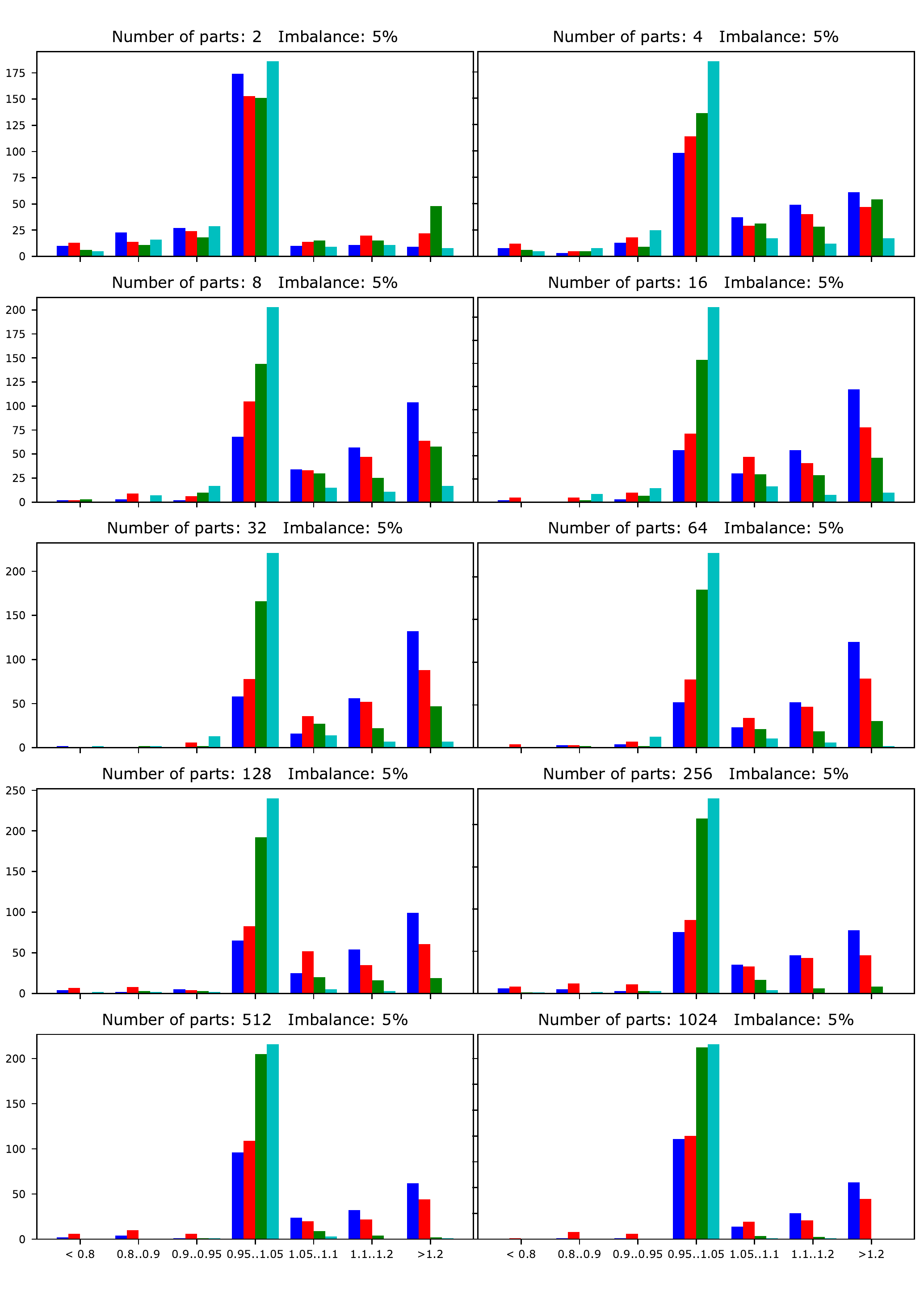}
  \caption{Histogram of $\zeta$ for coarsening using stable matching aggregation with imbalance factor 5\%. Blue rectangle corresponds to PaToH, red to hMetis2, green to Zoltan PHG and cyan to Zoltan-AlgD}\label{fig:stbres105}
\end{figure}

Figures~\ref{fig:aggres103}, \ref{fig:stbres103}, \ref{fig:aggres105} and \ref{fig:stbres105} present additional experimental results comparing the proposed schemes with state-of-the-art hypergraph partitioners. Figures~\ref{fig:aggres103} and \ref{fig:stbres103} present the results for imbalance factor 3\% and Figures \ref{fig:aggres105} and \ref{fig:stbres105} for imbalance factor 5\%.

\begin{figure}
\centering
  \includegraphics[width=\textwidth]{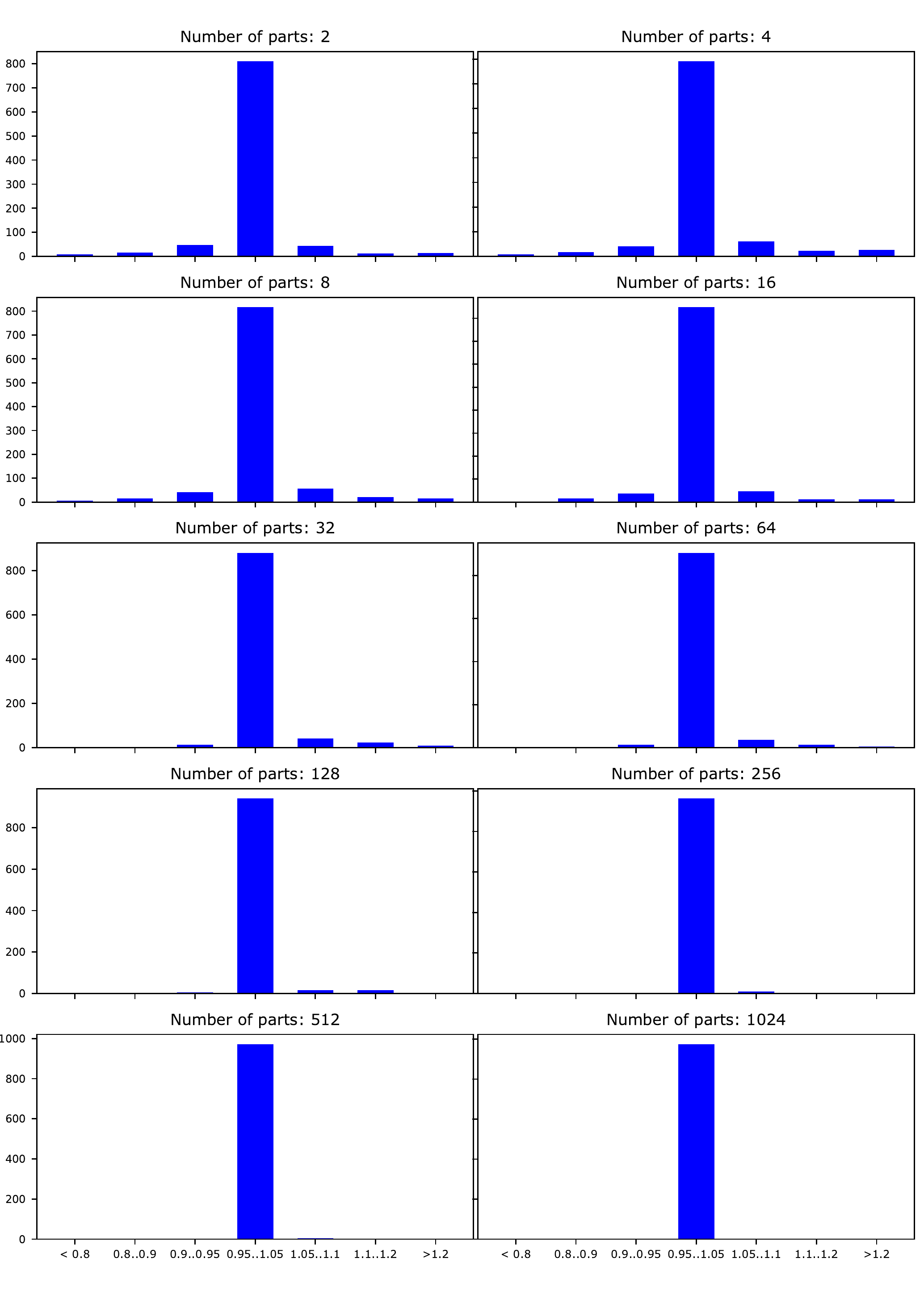}
  \caption{Histogram of $\zeta$ comparing coarsening using inner-product aggregation (denominator of $\zeta$) with coarsening using stable matching aggregation (nominator of $\zeta$) with imbalance factor 10\%. It is easy to see that the two algorithms perform very similarly.}\label{fig:stb_vs_agg}
\end{figure}

Figure~\ref{fig:stb_vs_agg} presents comparison of the quality of the solutions for the two proposed coarsening methods. Analogously to other barcharts, this figure compares the two algorithms by presenting the value

\begin{equation}
\zeta = \frac{\text{cut obtained using stable matching aggregation}}{\text{cut obtained using inner-product aggregation}}.
\end{equation}

The figure demonstrates that two algorithms produce solutions of very similar quality.

\section{Interesting observation about algorithmic variations and additional parameters for algebraic multigrid inner-product aggregation}\label{app:comparison}

\begin{figure}
\centering
  \includegraphics[width=\textwidth]{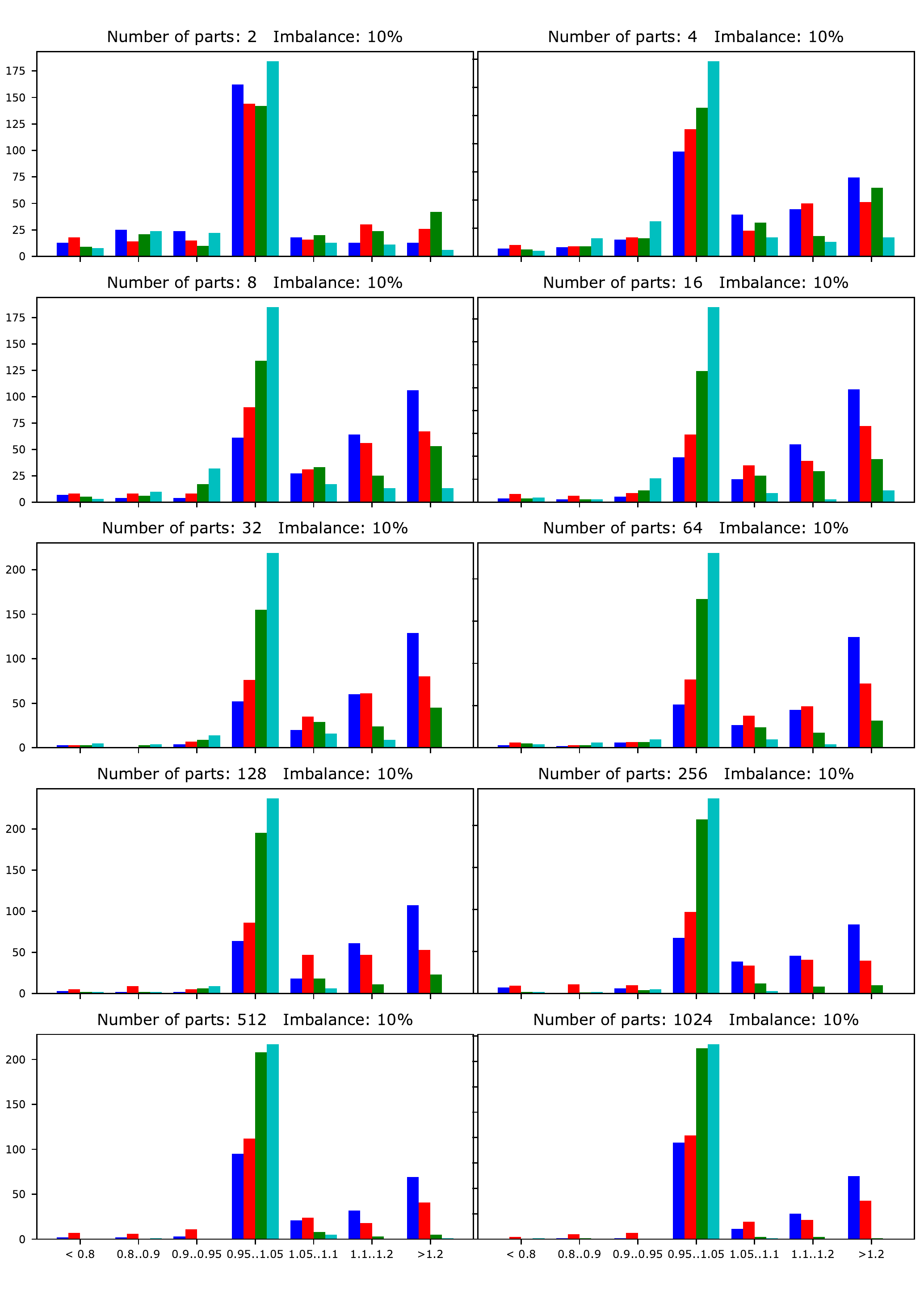}
  \caption{Histogram of $\zeta$ for coarsening using inner-product aggregation with additional parameters. Vertices are visited in the order of decreasing future volumes, inner-product metric is used. Blue rectangle corresponds to PaToH, red to hMetis2, green to Zoltan PHG and cyan to Zoltan-AlgD}\label{fig:addresfsort_no_norm}
\end{figure}

\begin{figure}
\centering
  \includegraphics[width=\textwidth]{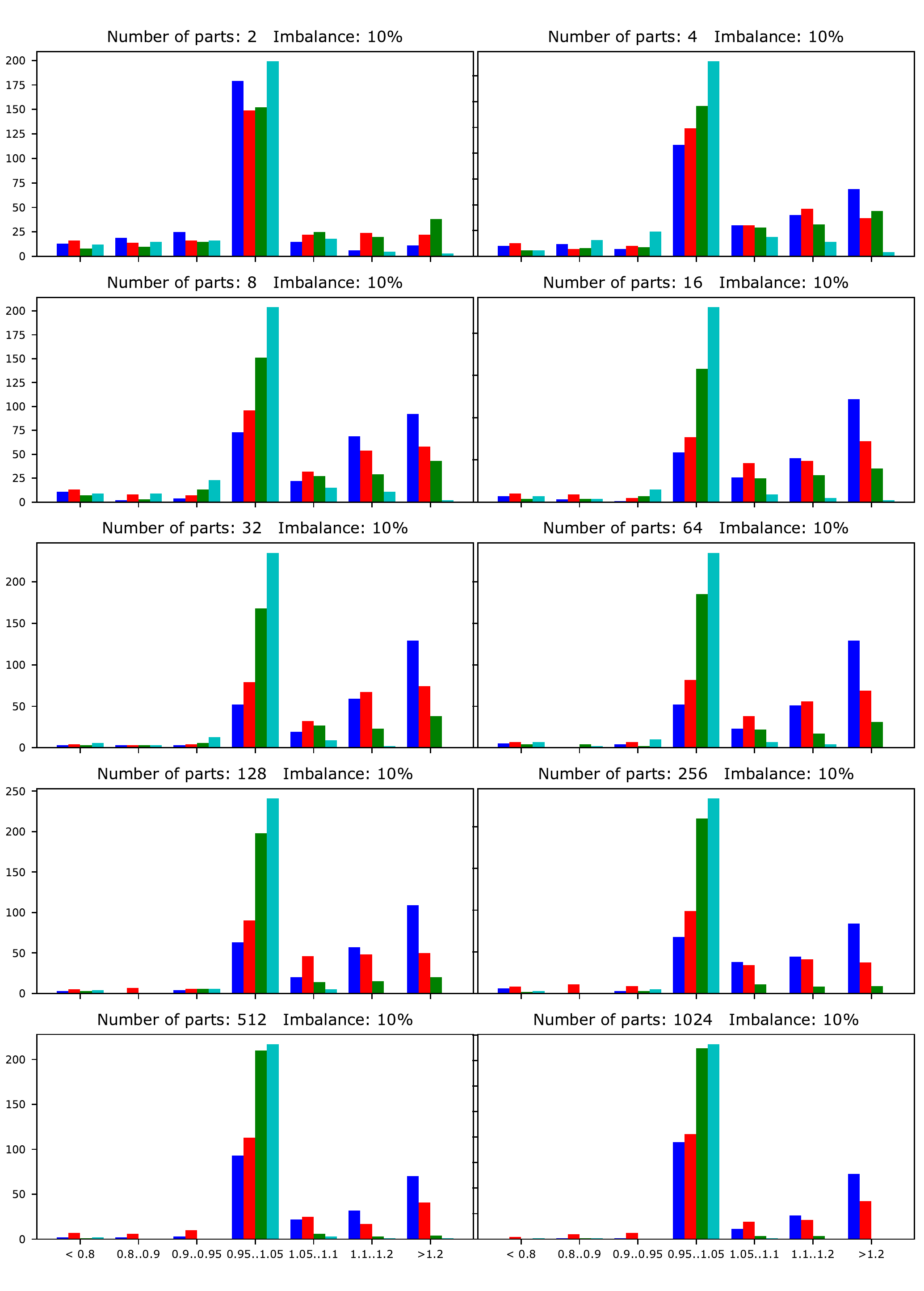}
  \caption{Histogram of $\zeta$ for coarsening using inner-product aggregation with additional parameters. Vertices are visited in the order of decreasing future volumes, connectivity metric is used. Blue rectangle corresponds to PaToH, red to hMetis2, green to Zoltan PHG and cyan to Zoltan-AlgD}\label{fig:addresfsort_norm}
\end{figure}

\begin{figure}
\centering
  \includegraphics[width=\textwidth]{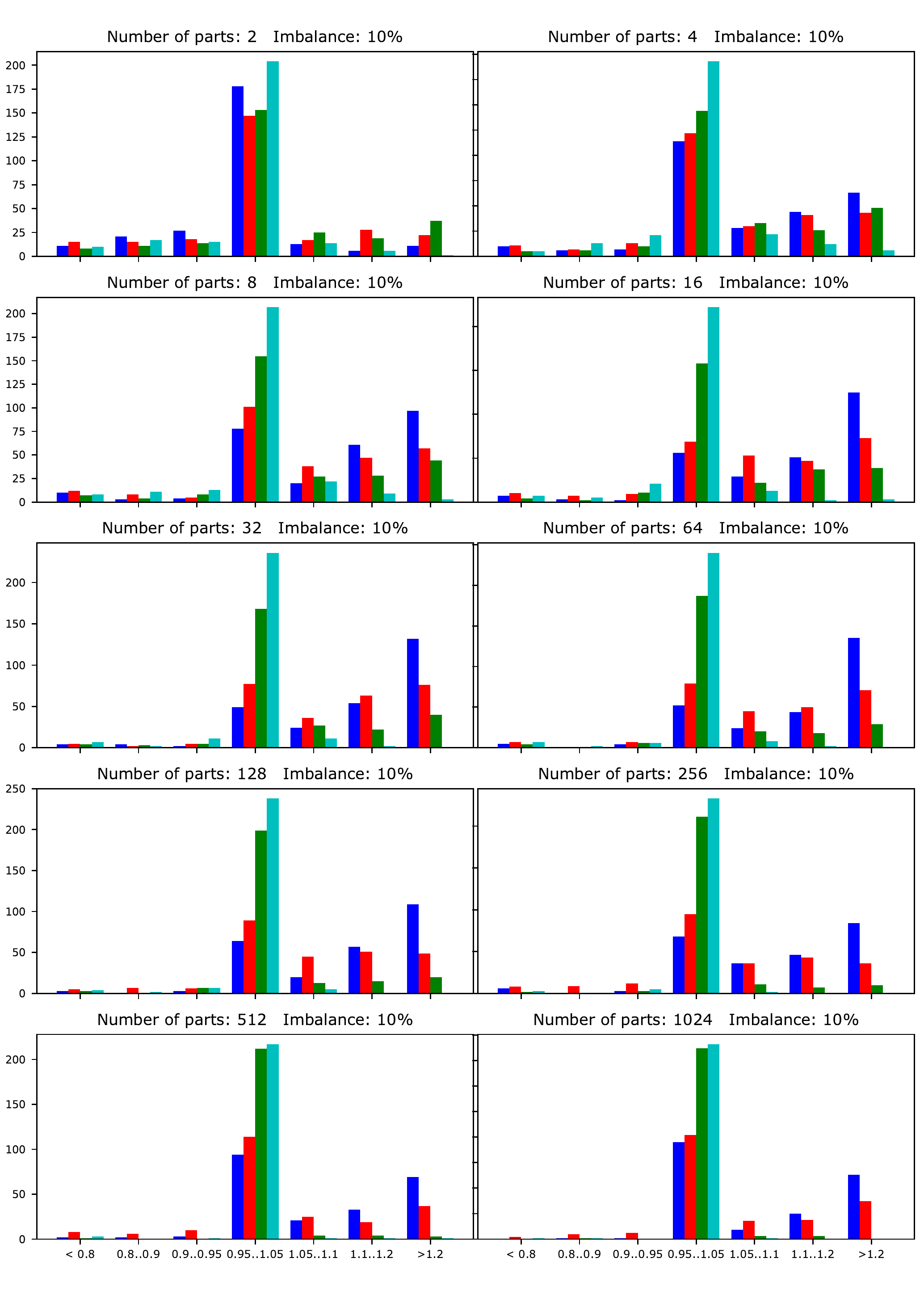}
  \caption{Histogram of $\zeta$ for coarsening using inner-product aggregation with additional parameters. Vertices are visited in random order, connectivity metric is used. Blue rectangle corresponds to PaToH, red to hMetis2, green to Zoltan PHG and cyan to Zoltan-AlgD}\label{fig:addresfno_sort_norm}
\end{figure}

We explore two additional approaches to inner-product aggregation (see Listing~\ref{code:ipm_match}). First, instead of visiting the vertices in random order, we investigate visiting them in the order of decreasing future volume. Second, instead of using the inner-product as a metric when selecting a seed to join, we explore using the connectivity metric: each non-seed $v$ a neighboring seed $u$ with the highest connectivity is selected and $v$ is added to the cluster $C_u$ seeded by it. The connectivity is defined as $N_{v,C_u}/W_{v,C_u}$, where $N_{v,C_u}=\Sigma_{e \mid C_u\cap e \neq\emptyset, v\in e}\ \rho(e)$ is the total algebraic weight of the edges connecting $v$ with the vertices in the cluster $C_u$, and $W_{v,C_u}=\Sigma_{i\in C_u} w(i) + w(v)$ is the total weight of the vertices in the potential cluster~\cite{catalyurek1999hypergraph}. The results of combinations of these two variations are presented in Figures~\ref{fig:addresfsort_no_norm}, \ref{fig:addresfsort_norm} and \ref{fig:addresfno_sort_norm}. All include imbalance factors of 10\%.

Surprisingly for us, we observed relatively insignificant differences in the results of these two (on the first glance) very important variations. The variation related to the strength of connectivity that depends on the capacity of already chosen cluster directly affects the size of the coarse aggregate. Too big aggregates can result in additional work (and thus computational time) of the refinement and getting trapped in false local attraction basins with KL/FM refinement. However, we observe that the entire framework resolves this issue without any problems.

%%
%% Bibliography
%%

%% Either use bibtex (recommended), 

\newpage 

\bibliography{fullbib,lipics-v2016-sample-article}

%% .. or use the thebibliography environment explicitely

\end{document}